\documentclass[twocolumn,showpacs,preprintnumbers,amsmath,amssymb]{revtex4}
\usepackage{amsmath}
\usepackage{graphicx}
\begin{document}

\title{\bf Theory of optical conductivity in detwinned YBa$_2$Cu$_3$O$_{y}$

}

\author
{
 T. Zhou$^{1,2}$, Jian-Xin Li$^{1,3}$, and Z. D. Wang$^{2,1}$
}

\address{
$^{1}$National Laboratory of Solid State Microstructures and
Department of Physics, Nanjing University, Nanjing 210093, China\\
$^{2}$Department of Physics and Center of Theoretical and
Computational Physics, University of Hong Kong, Pokfulam
Road, Hong Kong, China\\
$^{3}$ Department of Physics,University of California at Berkeley,
Berkeley, CA 94720, USA }

\date{\today}

\begin{abstract}
We study theoretically the chain and in-plane optical
conductivities of detwinned YBa$_2$Cu$_3$O$_{y}$. We elaborate
that the chain is superconducting for $y>6.67$
while  insulating for $y<6.67$ due to
the competition between the plane-chain coupling and the
antiferromagnetic order, corresponding to a superconductor-insulator
transition. Stemming also from  the coupling between the plane and
chain, a new peak emerges at a low frequency in the in-plane spectra
in the superconducting state, while it  disappears in the normal
state. Our scenario accounts satisfactorily for very recent
experiments.

\end{abstract}

 \maketitle

YBa$_2$Cu$_3$O$_y$ (YBCO) is one of the most studied high-$T_c$
superconducting materials. A primary distinction between this system
and other cuprates is the presence of the quasi-one-dimensional CuO
chain. Though it is widely believed that the main physics in this
material is within the CuO$_{2}$ plane, the study of electronic
properties in the chain and their influence on the superconducting
plane may enable us to have a profound understanding on
superconductivity in the present system~\cite{edw,der,lud,gag}. One
of powerful tools to detect the electronic structure is the
measurement of optical conductivity. It was reported that a
pronounced $a-b$ axis anisotropy exists in the
spectra~\cite{sch,rot,sch1,coo,shi,bas,lee}, which are strongly
enhanced in the chain direction. By subtracting the $a$-axis spectra
from the $b$-axis (the chain direction) spectra, one can obtain
experimentally the spectra contributed by the chain. In this way,
the chain electronic structure revealed by the optical experiments
shows some unusual features. Several experiments showed that the
optical conductivity in the CuO chain exhibits a peak at a finite
frequency~\cite{sch,rot,sch1} and approaches to zero in the dc
limit, implying that the chain is insulating though the CuO$_{2}$
plane is still in the superconducting state. On the other hand, it
was also reported that the chain spectra are characterized as a
Drude-like peak located at the zero frequency, demonstrating the
chain superconductivity~\cite{coo,shi,bas}.

Recently, a detail investigation of the infrared response in the
detwinned YBCO~\cite{lee} provided us a systematic doping
dependence of the electrodynamics in the CuO chain. The optical
spectra contributed by the chain display a dominant Drude-like
peak in the $y=6.75$ sample. While as $y$ decreases to 6.65, the
Drude-like peak will shift to a narrow resonance peak at the
finite frequency, indicating that the superconductor-insulator
transition occurs in the CuO chain. In fact, the appearance of
this transition around $y=6.65$ may also be inferred from the
doping dependence of the superfluid density~\cite{bas}. A
theoretical explanation of the chain superconductivity based on
the plane-chain coupling was given by Morr and
Balatsky~\cite{morr}, however, an elaboration of the observed
superconductor-insulator phase transition and the insulating
nature of the chain is still awaited.
In the meantime, another important issue is how the in-plane
spectra are influenced by the chain. Very recently, an experiment
on detwinned ortho-II phase YBa$_2$Cu$_3$O$_{6.5}$ observed an
extra strong peak at 180 cm$^{-1}$ (about 20 meV) in the $a$-axis
optical conductivity in addition to the usually observed
Drude-like peak and the mid-infrared (MIR) component in twinned
samples~\cite{hwa}. Therefore, a consistent accounting for the
optical response in both the plane and chain of the detwinned
system is appealing.


In this paper, we demonstrate that the competition of the
plane-chain coupling and the antiferromagnetic (AF) order in the
chain gives rise to the superconductor-insulator transition in the
CuO chain, namely, the proximity superconductivity in the chain is
induced by the plane-chain coupling when there is no AF order or
it is negligible, while the insulating gap in the optical
conductivity is caused by the AF order when its magnitude is
appreciable. Starting from a self-consistent mean-field treatment
for both the planar and chain Hamiltonian, we extend a simple
existing model~\cite{zho,zho1} of the plane-chain coupling to
include the emergency of the AF order in the chain at low dopings.
Interestingly, we find a step-like rise of the AF order at the
oxygen content $y=y_{c}=6.67$ which is almost the same as
the superconductor-insulator transition point observed
experimentally~\cite{lee}. In the meantime, the chain optical
conductivity shows an insulating gap below $y_{c}$; while above
$y_{c}$, the chain optical conductivity rapidly acquires a
Drude-like peak at the dc limit and shows a proximity-induced
superconductivity. On the other hand, we find that a new peak
occurs at a low frequency ($0.18J\approx 23$meV with $J$ as the AF
exchange integral) in the in-plane optical conductivity due to the
coupling between the plane and chain. These results agree
quantitatively with the recent experimental
measurements~\cite{lee,hwa}, and thus give a consistent picture of
the chain and in-plane optical conductivity in the detwinned YBCO.


We start with a Hamiltonian which describes a system with a plane
and a chain per unit cell,
\begin{equation}
H=H_{p}+H_c+H_I,
\end{equation}
where $H_p$ describes the CuO$_2$ plane:
\begin{eqnarray}
H_{p}&=&-t_p\sum_{\langle ij\rangle,\sigma}
c^{p\dagger}_{i\sigma}c^p_{j\sigma}+h.c.-t^{'}_p\sum_{\langle
ij\rangle{'},\sigma}c^{p\dagger}_{i\sigma}c^p_{j\sigma}+h.c
\nonumber\\&&+J\sum_{\langle
ij\rangle}({\bf{S}}^p_i\cdot{\bf{S}}^p_{j}-\frac{1}{4}n^{p}_in^{p}_j),
\end{eqnarray}
 $H_c$ describes the $y$-direction CuO chain:
\begin{equation}
 H_c=-t_c\sum_{\langle
ij\rangle,\sigma}c^{c\dagger}_{i\sigma}c^c_{j\sigma}+h.c.+J_c\sum_{\langle
ij\rangle}({\bf{S}}^c_i\cdot{\bf{S}}^c_{j}-\frac{1}{4}n^c_in^c_j),
\end{equation}
and $H_I$ is the coupling between the plane and chain:
\begin{eqnarray}\label{hi}
H_I&=&-t_{\perp}\sum_{ij\sigma}(c^{p\dagger}_{i\sigma}c^c_{j\sigma}+h.c).
\end{eqnarray}
Here $\langle ij\rangle$ denotes the nearest-neighbor (NN) bond and
$\langle ij\rangle^{'}$ the next NN bond.

First, we use the slave-boson mean-field theory to decouple the
Hamiltonians (2) and (3). Then, the coupling of the planar
fermions to spin fluctuations are included via the random phase
approximation (RPA). In the slave-boson approach, the creation
operators of the planar and chain electrons
$c^{p(c)\dagger}_{i\sigma}$ are expressed by slave bosons
$b^{p(c)}_i$ carrying the charge and fermions $f^{p(c)}_{i\sigma}$
representing the spin,
$c^{p(c)\dagger}_{i\sigma}=f^{p(c)\dagger}_{i\sigma}b^{p(c)}_i$.
The mean-field order parameters are defined as
$\Delta^{p}_{ij}=\langle
f^p_{i\uparrow}f^p_{j\downarrow}-f^p_{i\downarrow}f^p_{j\uparrow}\rangle=\pm\Delta_p$,
($\pm$ depend on if the bond $\langle ij\rangle$ is in the
$\hat{x}$ or the $\hat{y}$ direction), $\chi^{p(c)}_{ij}=\langle
f^{p(c)\dagger}_{i\sigma}f^{p(c)}_{j\sigma}\rangle=\chi_{p(c)}$.
Notice that we do not assume any superconducting pairing in the
chain at the mean-field level. The AF order in the chain is
defined as $m_c=(-1)^i\langle S^{c}_{iz}\rangle$. At low
temperatures we are concerned with, the boson condensation is
assumed $b^{p(c)}_i\rightarrow\langle
b^{p(c)}_i\rangle=\sqrt{\delta_{p(c)}}$, where $\delta_{p(c)}$ is
the hole density in the plane (chain).

After Fourier transformation, the mean-field Hamiltonian of the
planar and chain fermions can be written as
\begin{eqnarray}
H_p&=&\sum_{{\bf k}\sigma}\varepsilon^p_{\bf k}f^{p\dagger}_{{\bf
k}\sigma}f^p_{{\bf k}\sigma}-\sum_{\bf k}\Delta^p_{\bf
k}(f^{p\dagger}_{{\bf k}\uparrow} f^{p\dagger}_{-{\bf
k}\downarrow}+h.c.)\nonumber\\&&+2N_pJ^{'}_p(\chi^{2}_p+\Delta^{2}_p),
\end{eqnarray}

\begin{eqnarray}
H_c&=&\sum_{{\bf k}\sigma}(\varepsilon^c_{\bf k}f^{c\dagger}_{{\bf
k}\sigma}f^c_{{\bf k}\sigma}+\varepsilon^c_{{\bf k}+{\bf
Q_c}}f^{c\dagger}_{{\bf k}+{\bf Q_c}\sigma}f^c_{{\bf k}+{\bf
Q_c}\sigma}) -2J_cm_c\nonumber\\&&\sum_{{\bf
k}\sigma}\sigma(f^{c\dagger}_{{\bf k}\sigma}f^c_{{\bf k}+{\bf
Q_c}\sigma}+h.c.)+2N_cJ_c(\chi^2_c+m^2_c),
\end{eqnarray}
where the summation over ${\bf k}$ in the chain Hamiltonian is in
the magnetic Brillouin zone (MBZ) due to the AF order, i.e.,
$-\pi/2<k_y\leq\pi/2$. ${\bf Q_c}$ is the one-dimensional AF wave
vector with ${\bf Q_c}=\pi/2$. $\varepsilon^{p(c)}_{\bf k}$ are
given by $\varepsilon^{p}_{\bf k}=-2(\delta_p t_p+J^{'}_p
\chi_p)(\cos k_x+\cos k_y)-4\delta_p t^{'}_p\cos k_x \cos k_y-\mu_p$
and $\varepsilon^{c}_{\bf k}=-2(\delta_c t_c+J_c \chi_c)\cos
k_y-\mu_c$, respectively. Correspondingly, the coupling of the
planar and chain fermions can be written as
\begin{equation}
H_I=\widetilde{t}_\perp\sum_{{\bf k}\sigma}(f^{p\dagger}_{{\bf
k}\sigma}f^c_{{\bf k}\sigma}+h.c.).
\end{equation}

The renormalized Green's functions for the planar fermions
$\hat{G}_p$ due to the coupling to spin fluctuations are calculated
by Dyson's equation in the Nambu representation:
\begin{equation}
\widehat{G}_p({\bf{k}},i\omega)^{-1}=\widehat{G}_{p0}({\bf{k}},i\omega)^{-1}-\widehat{\Sigma}_p({\bf{k}},i\omega),
\end{equation}
where the self-energy can be obtained from~\cite{zho},
\begin{eqnarray}
\widehat{\Sigma}_p({\bf{k}},i\omega)&=&\frac{1}{{\beta}N_{p}}\sum_q\sum_{i\omega_m}J^{2}({\bf
q})\chi_p({\bf q},i\omega_m)
\widehat{\sigma}_3\nonumber\\&&\widehat{G}_{p0}({\bf{k}}-{\bf{q}},i\omega-i\omega_m)\widehat{\sigma}_3,
\end{eqnarray}
$\widehat{G}_{p0}$ is the bare Green's function obtained from the
mean-filed Hamiltonian and $\chi_p({\bf
q},i\omega_m->\omega_m+i\delta)$ is the RPA-type  spin
susceptibility~\cite{zho}. The Green's function for the chain
fermions $\hat{G}_c$ in the Nambu representation is obtained
directly from the above mean-field Hamiltonian.

The real part of the optical conductivity $\sigma_1(\omega)$ is
given by
$\sigma_{1\alpha\alpha}(\omega)_{p(c)}=-\mathrm{Im}\Pi_{\alpha\alpha}(\omega)_{p(c)}/\omega$
($\alpha=\hat{x},\hat{y}$). Here the imaginary part of the
current-current correction function
Im$\Pi_{\alpha\alpha}(\omega)_{p(c)}$ is expressed as
\begin{eqnarray}
\mathrm{Im}\Pi_{\alpha\alpha}(\omega)_{p(c)}&=&\sum_{\bf k}\frac{\pi
e^{2}}{N_{p(c)}}\int \mathrm{d} \omega' [v_\alpha({\bf k})_{p
(c)}]^{2}[f(\omega+\omega')\nonumber
\\&&-f(\omega')]
\mathrm{Tr}[\hat{A}_{p(c)}({\bf
k},\omega+\omega')\hat{A}_{p(c)}({\bf k},\omega')]\nonumber,
\end{eqnarray}
where $v_\alpha$ is the $\alpha$-component of the quasiparticle
group velocity and $\hat{A}_{p(c)}({\bf k},\omega)$ is the spectral
function [$\hat{A}_{p(c)}({\bf
k},\omega)=-(1/\pi)$Im$\hat{G}_{p(c)}({\bf k},\omega)$].

\begin{figure}

\centering

\includegraphics[scale=0.5]{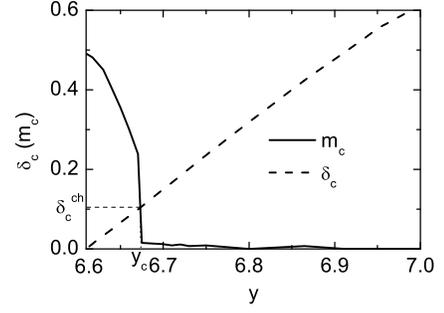}
\caption{The hole density $\delta_c$ (dashed line) and AF order
$m_c$ (solid line) in the chain versus the oxygen content $y$ in
YBa$_2$Cu$_3$O$_y$.} \label{fig1}
\end{figure}

In order to investigate the doping dependence of the optical
conductivity, we need to determine the planar and chain doping
density $\delta_{p(c)}$ for a given oxygen content $y$ in
YBa$_2$Cu$_3$O$_y$. We determine $\delta_{p}$ versus $y$ by using an
empirical relation proposed recently by Liang {\it et
al.}~\cite{liang}. On the other hand, in the parent compound
($y=6$), the valence of Cu ion in the CuO chain  is +1, and thus the
electron density in the chain is 2. 
Correspondingly, we may have a conservation condition
$2-n_{c}+2\delta_p=2(y-6)$, which relates the chain electron density
$n_c$ and the planar doping density $\delta_p$ with $y$. From this
condition, we can derive the chain doping density $\delta_c=1-n_c$
as a function of $y$, as shown in Fig.1. With the hole doping
density, we can then solve the set of self-consistent equations  for
mean-field parameters $\chi_{p(c)}$, $\Delta_{p}$, $m_{c}$, and
$\mu_{p(c)}$. The input parameters are, $t_c=t_p=2J$,
$t^{'}_p=-0.45t_p$, $J_c=J^{'}_p=3J/8$, $\widetilde{t}_\perp=0.1J$.
The AF coupling constant $J$ in the plane is used as the energy unit
($J\approx130$ meV). The chain AF order $m_c$ as a function of $y$
at temperature $T\approx0$ is shown as the solid line in Fig.1. We
can see that the chain AF order $m_c$ emerges with the decrease of
the chain doping density and reaches its maxima value at
$y=y_m=6.61$ where the chain doping density is
zero(half-filled)~\cite{note}. We note that the AF order has a
step-like rise from a negligible value to about a half of its maxima
at $y=y_c=6.67$.

The chain optical conductivity for different oxygen content $y$ is
plotted in Fig.2. As $y$ is increased to exceed  $y_c=6.67$
[Fig.2(a)],  a prominent Drude-like peak at the dc limit shows up
in the spectra, which indicates that the chain is superconducting.
When the oxygen content $y$ is decreased to  $y=6.67$, the
Drude-like peak disappears and a finite frequency peak at
$0.35J\approx 46$meV occurs, leaving an excitation gap between
this peak and the dc limit. This indicates that the system becomes
insulating at this doping density. With further decrease of the
oxygen content, the finite frequency peak will move to a higher
energy, leading to a larger excitation gap. Meanwhile, the
intensity of the spectra is suppressed heavily as shown in the
inset of Fig.2(b). So, the superconductor-insulator transition
occurs at $y=6.67$, and the corresponding critical doping density
in the chain is about $\delta^{ch}_c\approx 0.11$. This result is
consistent quantitatively with the recent experimental
data~\cite{lee}.

\begin{figure}
\centering
\includegraphics[scale=0.5]{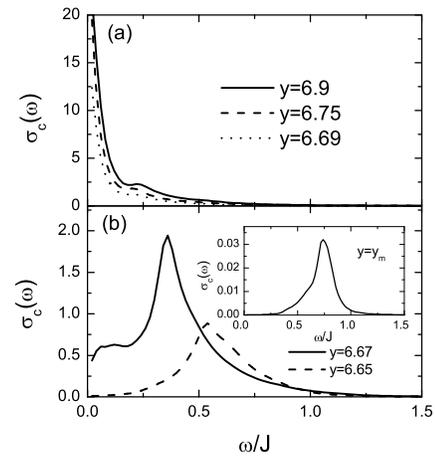}
\caption{The chain optical conductivity as a function of frequency
in YBa$_2$Cu$_3$O$_y$ for different oxygen content $y$. The inset
of Fig.(b) shows the spectra at $y=y_m=6.61$ where the calculated
chain doping density is zero(half-filled).} \label{fig2}
\end{figure}

 From Fig.1, one can see that the AF order is negligible or
disappears (for $y>6.9$) in the doping range where a Drude-like peak
in the optical spectra is observed and has an appreciable value when
the optical spectra show an insulating behavior. More
importantly, at 
$y=y_{c}=6.67$ where the insulator-superconductor transition occurs,
the magnitude of AF order has a step like increase, i.e., it rises
from a negligible value to nearly a half of its maximum. This shows
clearly that the transition is associated with the emergency of the
AF order. The occurrence of the AF order leads to a gap
$2J_cm_c\approx 24.4$meV in the single-particle energy band of the
chain, i.e., a 48.8meV excitation gap in the particle-hole
excitations. This excitation gap suppresses the proximity effect
coming from the coupling to the superconducting plane which is
otherwise effective for $y>y_{c}$. Therefore, it is the competition
between the AF order and the proximity effect that leads to the
superconductor-insulator transition in the chain.



When $m_c$ equals to zero, the Hamiltonian for the system can be
written as $4\times4$ matrix~\cite{zho,zho1}. The gap symmetry
induced by the proximity effect may be examined by looking at the
abnormal Green's functions of the chain electrons, $ F_c({\bf
k},\omega)=\sum_i U_{3i}({\bf k},\omega)U_{4i}({\bf
k},\omega)/[i\omega-E_i({\bf k},\omega)], $ with $U_{4j}({\bf
k},\omega)=\widetilde{\Delta}^p_{\bf
k}\widetilde{t}_\perp^{2}A_{j\bf k}/F_{j\bf k}$ and
$\widetilde{\Delta}^p_{\bf k}$ the superconducting gap in the plane
with a $d_{x^2-y^2}$ symmetry~\cite{zho}. Thus, the induced gap of
the chain fermions is also of $d_{x^2-y^2}$ symmetry.

Now we study how the in-plane optical conductivity is influenced by
the coupling to the chain. The real part of the in-plane optical
conductivity $\sigma_{1}(\omega)$ is shown in Fig.3 for the
superconducting [Fig.3(a-b]) and the normal state [Fig.3(c)],
respectively. In the superconducting state, a Drude-like peak at the
dc limit and a MIR hump around $\omega\sim J$ is evident from
Fig.3(a) and (b), which reproduces what has been observed in the
twinned samples~\cite{basov}. These features are also consistent
with previous theoretical calculations\cite{stojkvic,munzar,zha} in
which only the in-plane electrodynamics is considered. A new feature
observed here is that an extra peak emerges between the Drude-like
peak and the MIR hump. Moreover, this extra peak disappears
completely in the normal state, as seen from Fig.3(c). These results
are consistent with the very recent experimental data~\cite{hwa}.
We also note that there is only a weak $a-b$ anisotropy in the
spectra around this peak. In the meantime, the in-plane spectra in
the superconducting state are nearly isotropic both in the dc
limit and in high frequencies. As a result, the plane-chain
coupling does not cause an obvious in-plane anisotropy in the
optical conductivity. Therefore, we expect that the anisotropy
observed in experiments, which is obtained by subtracting the
$a$-axis spectra from the $b$-axis spectra, is mainly contributed
by the chain contribution.
\begin{figure}
\centering
\includegraphics[scale=0.4]{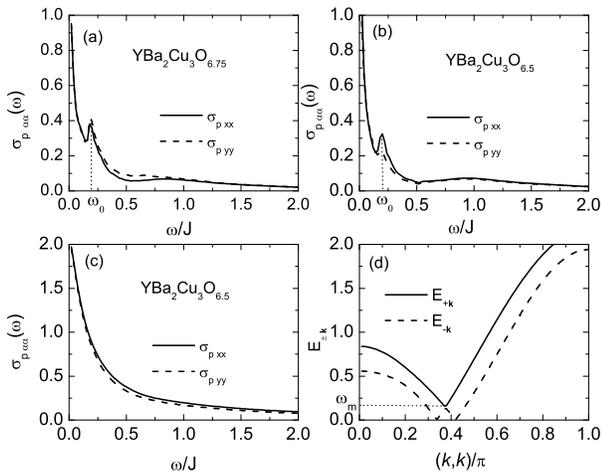}
\caption{Panels a)-c) plot the real parts of the in-plane optical
conductivity as a function of frequency $\omega$ for  different $y$.
Panel a) shows the spectra for $y=6.75$ sample in the
superconducting state, panels b) and c) are the spectra for the
ortho-II phase $y=6.5$ sample with the planar doping density
$\delta_p=0.097$~\cite{liang} in the superconducting state at
$T=0.0005J$ and in the normal state at $T=T_c=0.033J$, respectively.
Panel d) shows the quasiparticle energy $E_{\pm\bf k}$ along the
diagonal direction in YBa$_2$Cu$_3$O$_{6.75}$.} \label{fig3}
\end{figure}

Due to the coupling between the plane and chain, the energy band
of quasiparticle is split into two branches with frequencies
$E_{\pm{\bf k}}$~\cite{zho}. In the superconducting state, the
low-energy optical response comes mainly from the charge
excitations around the nodal direction due to the presence of the
superconducting gap. In Fig.3(d), we plot $E_{\pm{\bf k}}$ with
${\bf k}$ along the diagonal direction. Around the Fermi wave
vector, there is a minimum $\omega_{m}\approx 0.18J$ in $E_{+{\bf
k}}$, and the band $E_{-{\bf k}}$ is below this minimum. When the
excitation frequency is below $\omega_{m}$, the optical response
comes only from the $E_{-{\bf k}}$. When $\omega \ge \omega_{m}$,
an additional scattering channel from the band $E_{+{\bf k}}$
contributes to the response and leads to a peak around
$\omega_{m}$. In the normal state, the excitations near the entire
Fermi surface are available, so that the minimum in $E_+$ varies
at different wave vectors. As a result, the combined effect of
these excitations causes no extra peak.

To summarize, we have elaborated that the competition of the
plane-chain coupling and the antiferromagnetic order in the chain
leads to the superconductor-insulator transition in the optical
conductivity of the CuO chain. Specifically, the proximity
superconductivity in the chain is induced by the plane-chain
coupling while the insulating property is caused by the
antiferromagnetic order. Meanwhile, we find an extra peak in the
in-plane optical spectra at low frequencies in the superconducting
state, which is between the Drude-like peak and mid-infrared
component. This peak disappears in the normal state. These results
are well consistent with the recent optical conductivity
measurements.

The work was supported by the NSFC (10174019,10334090,10429401 and
10525415), the RGC grants of Hong Kong(HKU 7050/03P and HKU7045/04P)
and the 
973 project under the Grand No.2006CB601002. JXL acknowledges the
support of the Berkeley Scholar program.


\begin{thebibliography}{99}
\bibitem{note} The value of $y_m$ depends weekly on the relation $\delta_p\sim y$ because $\delta_p\ll 2-2(y-6)$ as $y=y_m$.
\end{thebibliography}
\end{document}